\documentclass[11pt,twoside]{article}
\usepackage{./asp2014}

\aspSuppressVolSlug
\resetcounters

\begin{document}

\title{Modeling the Complete Lightcurve of $\omega$ CMa}
\author{M. R. Ghoreyshi $^{1}$, A. C. Carciofi$^{1}$, L. R. R\'{\i}mulo$^1$, S. Otero$^2$, D. Baade$^3$, J. E. Bjorkman$^4$, A. T. Okazaki$^5$, and Th. Rivinius$^6$
\affil{$^1$Instituto de Astronomia, Geof\'{\i}sica e Ci\^{e}ncias Atmosf\'{e}ricas, Universidade de S\~{a}o Paulo, Rua do Mat\~{a}o 1226, S\~{a}o Paulo, SP 05508-900, Brazil; \email{mohammad@usp.br}}
\affil{$^2$American Association of Variable Star Observers (AAVSO), Cambdrige, MA, USA;}
\affil{$^3$European Organisation for Astronomical Research in the Southern Hemisphere, Karl-Schwarzschild-Str. 2, 85748 Garching bei M\"{u}nchen, Germany;}
\affil{$^4$EDepartment of Physics  Astronomy, University of Toledo, MS111 2801 West Bancroft Street, Toledo, OH 43606, USA;}
\affil{$^5$Faculty of Engineering, Hokkai-Gakuen University, Toyohira-ku, Sapporo 062-8605, Japan;}
\affil{European Organisation for Astronomical Research in the Southern Hemisphere, Santiago 19, Casilla 19001, Chile}}


\begin{abstract}
We have used the radiative transfer code {\sf HDUST} to analyze and interpret the long-term photometric behavior of the Be star $\omega$ CMa, considering four complete cycles of disk formation and dissipation. This is the first time in which a full lightcurve of a Be star was investigated and modeled including both disk build-up and dissipation phases. Based on the quite good fit of the observed data we were able to derive the history of stellar mass decretion rates (including long- and short-term changes) during the disk formation and dissipation phases in all four cycles.

\end{abstract}

\section{Introduction}

 $\omega$ (28) CMa (HD 56139, HR2749; B3Ve) is one of the most observed southern Be stars. Long-term photometric monitoring exhibits a quasi-regular cyclic variation with an amplitude of about 0.$^{m}$5 in the $V$-Band. This allows to study evolution of the disk in detail. In each cycle a new disk is formed during 2-3 years, and then dissipated in 5-6 years.

Carciofi et al. (2012) used $\omega$ CMa to model, for the first time, the light curve of a Be star based on the viscous decretion disk (VDD) model (the theoretical description of the model is given by Haubois et al. 2012). The model of the dissipation curve allowed the authors to determine that the viscosity parameter of Shakura-Sunyaev (Shakura \& Sunyaev, 1973) is $\alpha$ = 1.0 $\pm$ 0.2. Such value for $\alpha$ suggests that viscosity may be produced by a disk instability whose growth is limited by shock dissipation (Carciofi et al., 2012).

\articlefigure{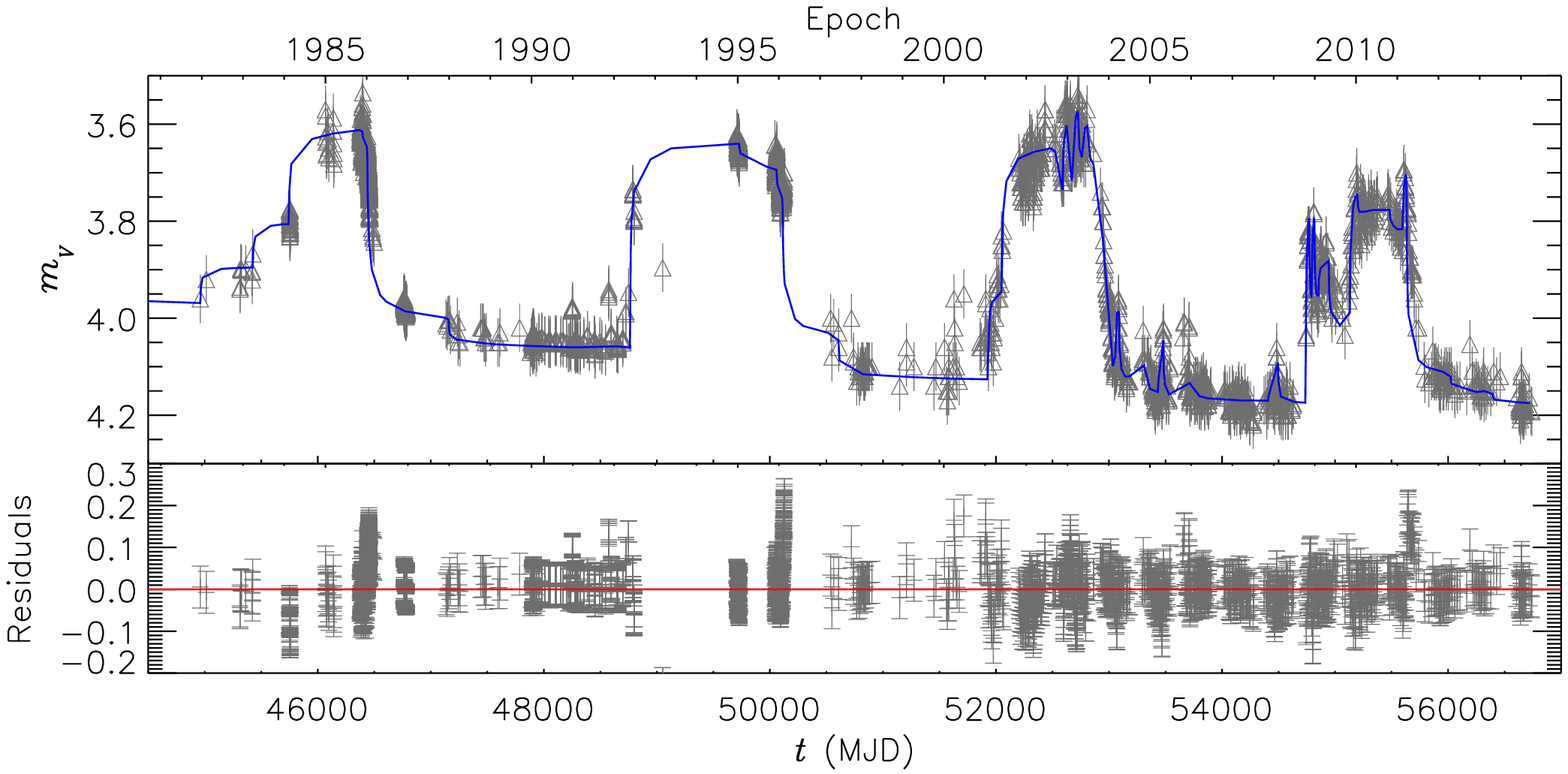}{fig1}{{\emph Top:} Observed light curve (grey triangles) of $\omega$ CMa compared with our best-fit theoretical model (blue line).
{\emph Bottom:}  Residuals of the fit.}

\articlefigure{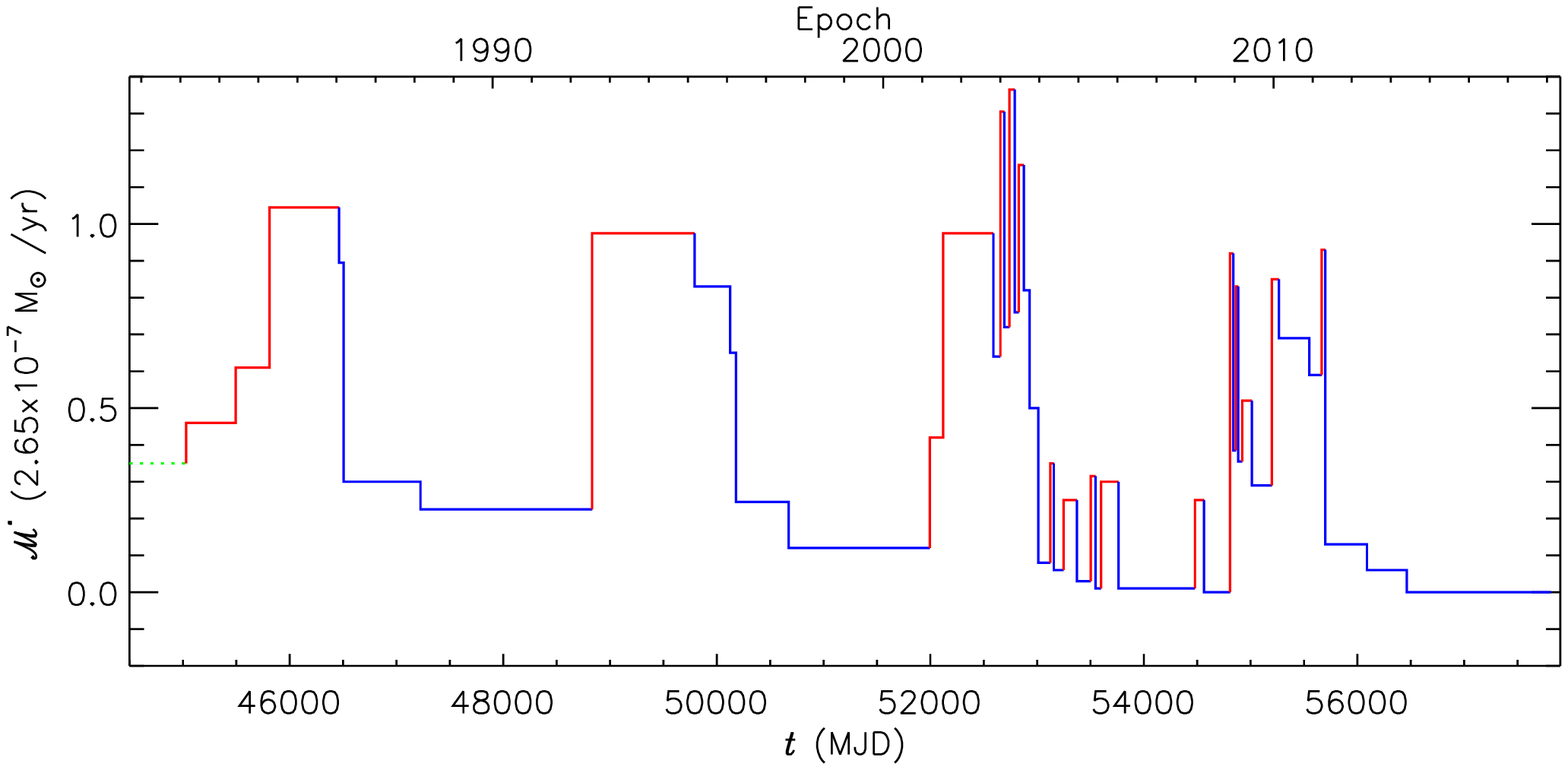}{fig2}{The mass decretion rate history of $\omega$ CMa: The red lines show the outbursts and the blue lines show the quiescence.}

\articlefigure{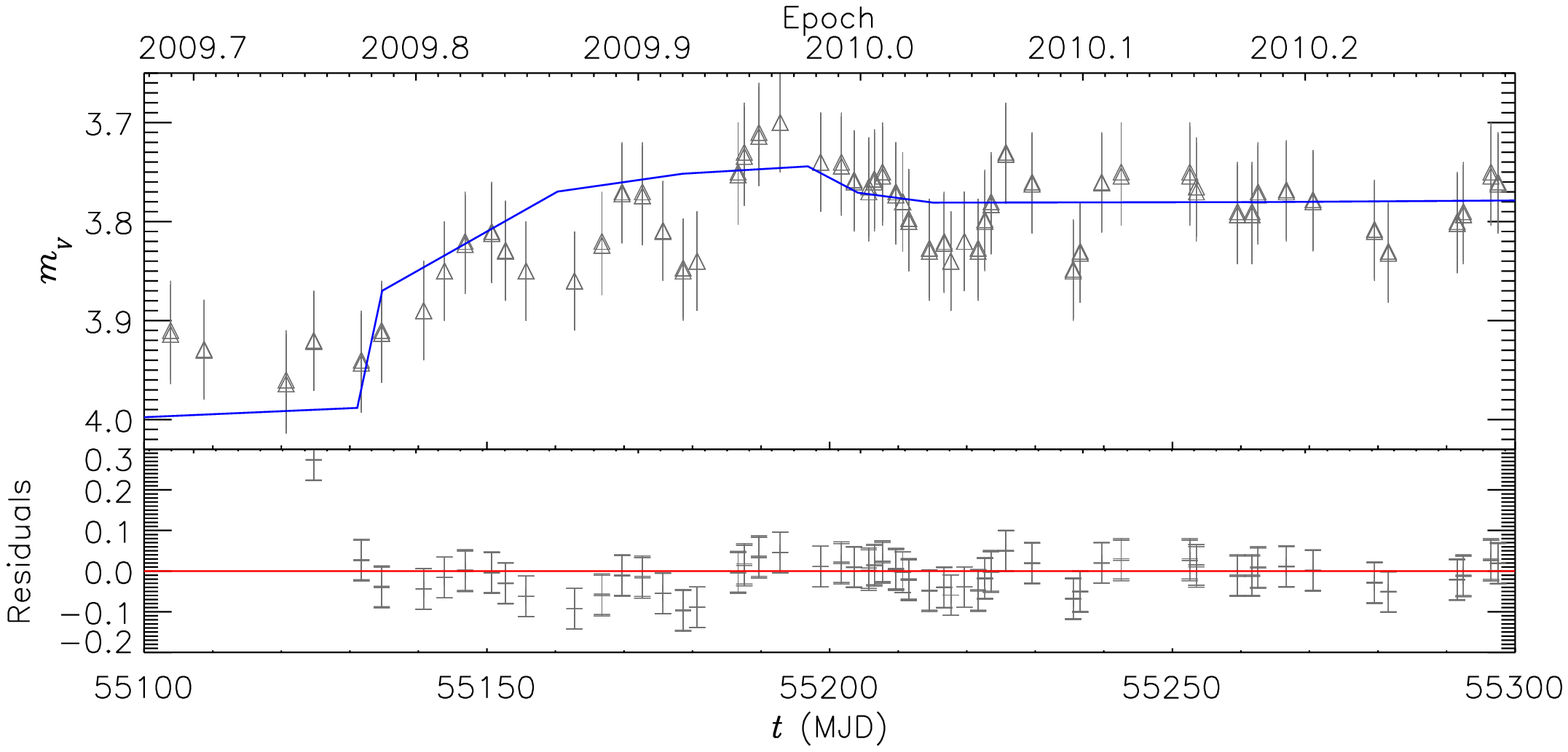}{fig3}{Consistency of model based on $\alpha$ =1.0 with the observed data even in the small events: This small outburst (phase $I$ in Fig. 4) happened in the middle of main outburst phase of fourth cycle, after a small quiescence. Theoretical lightcurve is shown by blue solid line and the observed data is shown by the grey triangles.}

\articlefigure{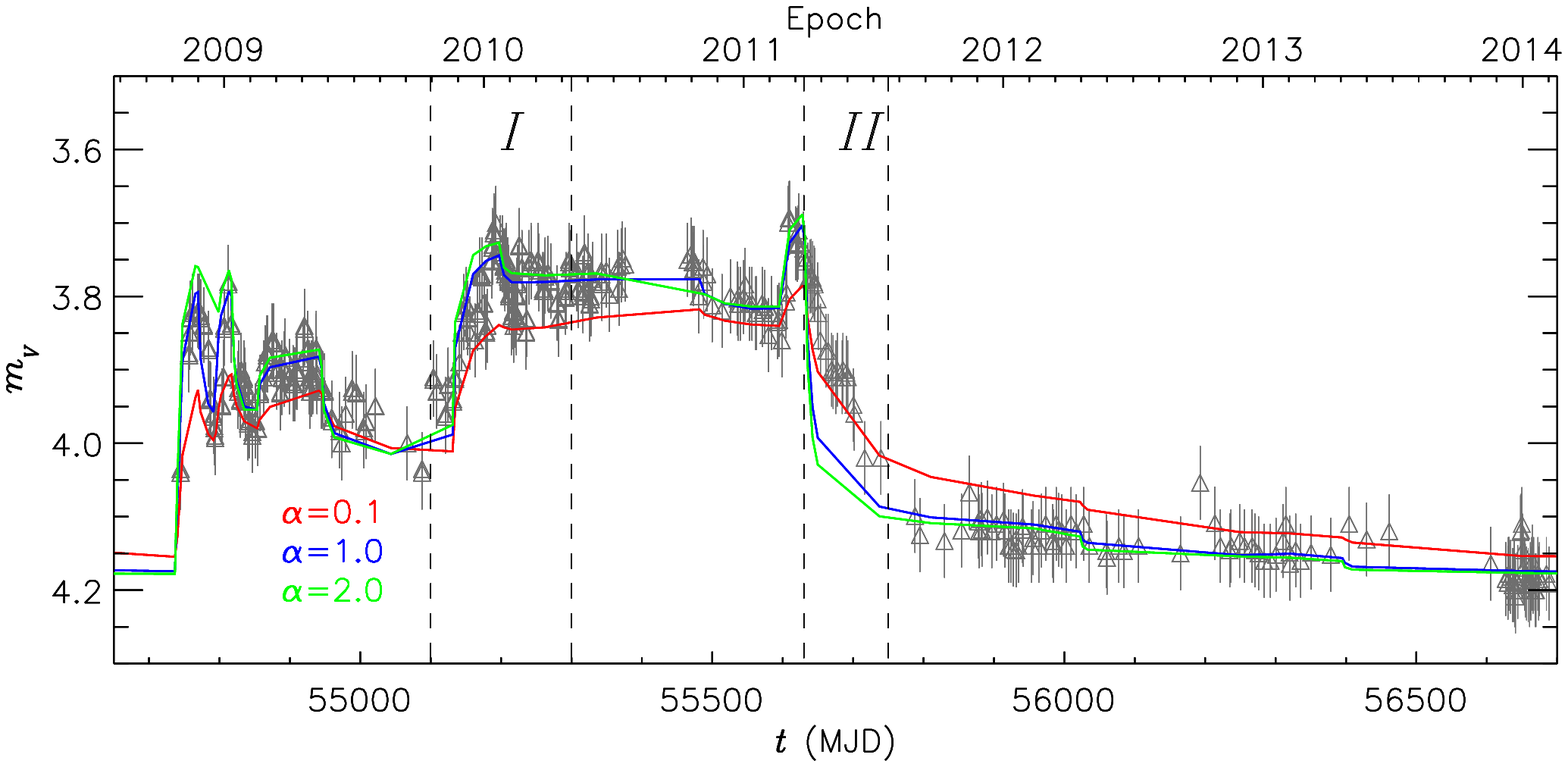}{fig4}{Theoretical lightcurve of $\omega$ CMa for different viscosity parameters: $\alpha$ =0.1 (red solid line), 1.0 (blue solid line, 2.0 (green solid line: The lightcurve for $\alpha$ =1.0 has broad consistency with the observed data, but does not fit well on the last rapid decline (Phase $II$).}

The work of Carciofi et al. (2012) represented just a first step towards understating the physical conditions of $\omega$ CMa's disk. As such, the study had several limitations, among which we cite the fact that $\alpha$ was determined only during the phase of disk dissipation and by fitting only the $V$-band light curve. Lifting these limitation is the main motivation for this work.

\section{Results}
 We used the time-dependent hydrodynamics code {\sf SINGLEBE} (Okazaki et al. 2002; Okazaki 2007) to model the disk surface density evolution given a prescription for the disk feeding history. To do that, several outbursts and quiescence phases with various time-scales, few weeks to few years, were considered. The surface density converted to volume density then is used as input for the three-dimensional Monte Carlo radiative transfer code, {\sf HDUST} (Carciofi \& Bjorkman 2006) to obtain the emergent spectral energy distribution (and other observables of interest), which is used to calculate the $V$-band excess, $\Delta$$V$ , of the disk as a function of time.

For the results of Fig. 1 we adopted as stellar parameters the ones used by Carciofi et al. (2012) and let the mass decretion rate vary in amplitude with time, in a fashion described in Fig. 2.

One of the features of $\omega$ CMa's lightcurve is an obvious decline in the brightness of the system in the successive dissipation phases (green line in Fig. 1). We found that one possibility to interpret this behavior is to consider that at quiescence the star shifts to a lower mass decretion rate instead of entering a state of absolute quiescence.

Our model shows the value of $\alpha$ should not be far from 1.0 at least for the most part. In Fig. 3 we show the first model of a Be disk build-up (for phase $I$ in Fig. 4), and this model is broadly consistent with a alpha of unity.

The dissipation phase marked as II in Fig. 4 presented a challenge for the model, as it is seemingly not consistent with and alpha of unity. Instead, a much lower alpha=0.1 seems to better reproduce the data. This fit is still under investigation, since the slower dissipation in phase II may simply be the result of small-scale outbursts during the disk dissipation phase.

According to the obtained results, we found that the mass decretion rate of $\omega$ CMa displays a complex history including both short and long time-scales, also very different range of variation (Fig. 2).

\section{Discussion and Perspectives}

We present a complete model of the disk evolution of $\omega$ CMa for the past forty years. For the first time, both outburst and quiescence phases were modeled with a single model.

The overall fit is quite good, showing strong evidence of the VDD scenario as the mechanism controling the disk formation and evolution.

A very puzzling feature of $\omega$ CMa is the slow secular fading shown in the past forty years, in which each quiescent phase is less bright than the previous one.  Here, we present one possible explanation for this phenomenon, by assuming that the mass decretion rate did not go to zero at quiescence, but instead assumed a smaller value (Fig. 2). The validity of this assumption needs to be ascertained by modeling other observables in addition to $V$-band magnitude. This will be the subject of future work.

\acknowledgements
This work made use of the computing facilities of the Laboratory of Astroinformatics (IAG/USP, NAT/Unicsul), whose purchase was made possible by the Brazilian agency FAPESP (grant No 2009/54006-4) and the INCT-A. A.C.C acknowledges support from CNPq grant No 307076/2012-1.


\begin{thebibliography}{}
\bibitem[Carciofi (2006)]{ref_1}
Carciofi, A.C., Bjorkman, J.E., 2006, ApJ, 639, 1081
\bibitem[Carciofi (2012)]{ref_2}
Carciofi, A.C., Bjorkman, J.E., Otero, S.A., Okazaki, A.T., \v{S}tefl,S., Rivinius, Th., Baade, D., Haubois, X., 2012, ApJ, 744, 15
\bibitem[Haubois (2012)]{ref_3}
Haubois, X., Carciofi, A. C., Rivinius, Th., Okazaki, A. T., Bjorkman, J. E., 2012, ApJ, 756, 156
\bibitem[Okazaki (2002)]{ref_4}
Okazaki, A.T., Bate, M.R., Ogilvie, G.I., Pringle, J.E. 2002, MNRAS, 337, 967
\bibitem[Okazaki (2007)]{ref_5}
Okazaki, A. T., 2007, ASP Conference Series, 361, 230
\bibitem[Shakura (1973)]{ref_6}
Shakura, N. I., Sunyaev, R. A., 1973, A\&A, 24, 337
\end{thebibliography}


\end{document}